\documentclass[aps,prb,twocolumn,showpacs,showkeys,superscriptaddress,citeautoscript]{revtex4}
\usepackage{amsmath,amssymb,graphicx,graphics,color}
\usepackage{dcolumn}
\usepackage{bm}

\usepackage{amsfonts}
\usepackage{amssymb,amsmath}
\usepackage{graphicx}
\usepackage{dcolumn}
\usepackage{bm}
\usepackage{color}
\def\2{\frac{1}{2}} \def\4{\frac{1}{4}}
\def\6{\partial}
\def\+{\dagger}
\def\<{\langle} \def\>{\rangle}

\begin{document}
\title{Magnetization non-rational quasi-plateau and spatially modulated spin order in the model of the single-chain magnet,
$[\{(\text{CuL})_2\text{Dy}\}\{\text{Mo}(\text{CN})_8\}]\cdot\text{2CH}_3\text{CN}\cdot\text{H}_2\text{O}$}
\author{Stefano Bellucci}
\affiliation{INFN-Laboratori Nazionali di Frascati, Via E. Fermi 40,
00044 Frascati, Italy}
\author{Vadim Ohanyan}
\affiliation{Institut f\"{u}r Theoretische Physik, Leibniz Universit\"{a}t Hannover, 30167 Hannover, Germany}
\affiliation{Department of Theoretical Physics,
             Yerevan State University,
             Alex Manoogian 1, 0025 Yerevan, Armenia}
\affiliation{ICTP, Strada Costiera 11, I-34151 Trieste, Italy}
\author{Onofre Rojas}
\affiliation{Departamento de Ciencias Exatas, Universidade Federal
de Lavras, CP 3037, 37200000, Lavras, MG, Brazil}

\affiliation{ICTP, Strada Costiera 11, I-34151 Trieste, Italy}

\date{\today}

\begin{abstract}
Using the exact solution in terms of the generalized classical
transfer matrix method we presented a detailed analysis of the
magnetic properties and ground state structure of the simplified model of the single-chain
magnet, trimetallic coordination polymer compound,
$[\{(\text{CuL})_2\text{Dy}\}\{\text{Mo}(\text{CN})_8\}]\cdot\text{2CH}_3\text{CN}\cdot\text{H}_2\text{O}$,
in which L$^{2-}$ is
N,N'-propylenebis(3-methoxysalicylideneiminato). Due to appearance
of highly anisotropic Dy$^{3+}$ ion this material is an unique
example of the one-dimensional magnets with Ising and Heisenberg
bonds, allowing exact statistical-mechanical treatment. We found two
zero-temperature ground states corresponding to different part of
the magnetization curve of the material. The zero-filed ground state
is shown to be an antiferromagnetic configuration with spatial
modulation of the local Dy$^{3+}$(which is proven to posses well
defined Ising like properties due two large anisotropy of g-factors)
 and composite $S=1/2$ spin of the quantum spin trimer Cu-Mo-Cu in
 the form "up"-"down"-"down"-"up". Another important feature of this
 compound is the appearance of the quasi-plateau at non rational
 value of magnetization due to difference of the g-factors of the
 Cu- and Mo-ion in quantum spin trimers. The quasi-plateau is an
 almost horizontal part of the magnetization curve where the
 corresponding zero-temperature ground state of the chain
 demonstrate slow but monotonous dependence of the magnetization on
 the external magnetic field, while the $z$-projection of the total spin, $S_{tot}^z$ is constant.
\end{abstract}



\maketitle


Few years ago a peculiar compound belonging to the specific family
of magnetic materials, the single chain magnet (SCM) with Heisenberg
and Ising bonds arranged in a periodic way (Heisenebrg--Ising chain
(HIC)), has been synthesized\cite{Dy1,Dy2}. The special arrangement
of the interaction bonds between local magnetic moments makes the
system exactly solvable by means of the generalized classical
transfer
 matrix (GCTM) method
\cite{Val,ant09,oha09,oha10,bel10,roj11a,RS11,SSR11,str12,oha12,bor_un,bel12,PSR,RSL}.
However, the pioneering works on HIC dealt with another method, the
so-called decoration iteration
transformation\cite{str05,str03,can06}. It is worth mentioning, that
such systems have been the subject of quite intensive theoretical
investigations during last decade but mainly in the context of the
exactly solvable strongly interacting models of statistical
mechanics as well as a possible approximation for the purely quantum
spin-chains\cite{SSR11,str12,oha12,str05}. However, the discovery of
the SCM among the family of the Dy$^{3+}$ ion based coordination
trimetallic polymers\cite{Dy1} offers an exceptional opportunity of
application of the exactly solvable theoretical models of the HIC to
the thermodynamics of the real magnetic materials. In Ref.
[\onlinecite{Dy2}], with the aid of the GCTM technique for HIC, the
theoretical description of the magnetic properties on SCM,
$[\{(\text{CuL})_2\text{Dy}\}\{\text{Mo}(\text{CN})_8\}]\cdot\text{2CH}_3\text{CN}\cdot\text{H}_2\text{O}$,
in which L$^{2-}$ is
N,N'-propylenebis(3-methoxysalicylideneiminato), denoting for the
sake of simplicity by [DyCuMoCu]$_{\infty}$, has been performed. The
structure of its magnetic lattice has been presented in Fig.
\ref{fig1}. The corresponding Hamiltonian is a sum of mutually
commuting block Hamiltonians:

\begin{figure}[h]
\includegraphics[width=\columnwidth]{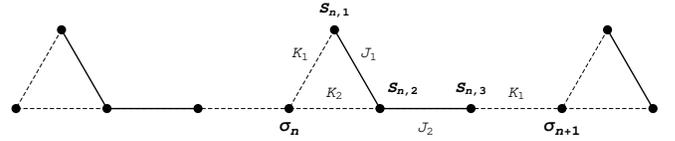}
\caption{\label{fig1} The magnetic structure of the single chain
magnet with Ising and Heisenberg bonds [DyCuMoCu]$_{\infty}$. Dashed
lines denote Ising couplings while solid lines correspond to
isotropic Heisenberg interactions. Spins $\mathbf{S}_{n,1}$ and
$\mathbf{S}_{n,3}$ correspond to Cu$^{2+}$ ions with $S$=1/2 and
isotropic\cite{footnote1} $g=2.16$, $\mathbf{S}_{n,2}$ is the spin of Mo$^{5+}$
S=1/2 ion with $g=1.97$, the $\sigma$-spins correspond to the
Dy$^{3+}$ highly anisotropic ions with $g_z=19.6$ and $g_x=g_y$=0.}
\end{figure}
\begin{eqnarray}\label{ham}
&&\mathcal{H}=\sum_{n=1}^N\left(\mathcal{H}_n- \mu_Bg_3\frac{B \cos \theta
}{2}\left(\sigma_n+\sigma_{n+1}\right)\right),\\
&&\mathcal{H}_n=J_1\mathbf{S}_{n,1}\cdot\mathbf{S}_{n,2}+J_2\mathbf{S}_{n,2}\cdot\mathbf{S}_{n,3}+K_2S_{n,2}^z\sigma_n\nonumber\\
&&+K_1\left(
S_{n,1}^z\sigma_n+S_{n,3}^z\sigma_{n+1}\right)\nonumber\\
&&-\mu_B g_1B\left(S_{n,1}^z+S_{n,3}^z\right)- \mu_B g_2B S_{n,2}^z,\nonumber
\end{eqnarray}
here the Heisenberg interaction between Cu$^{2+}$ ions with $S$=1/2
($\mathbf{S}_{n,1}$ and $\mathbf{S}_{n,3}$) and  Mo$^{5+}$ S=1/2 ion
($\mathbf{S}_{n,2}$) is supposed to be isotropic\cite{footnote1}, the highly
anisotropic Dy$^{3+}$ ions are denoted by the Ising variables
$\sigma_n$ taking $\pm 1/2$ values and the $g$-factors for all three
magnetic ions presented in the compound are different. As the magnetic measurements of the 
 [DyCuMoCu]$_{\infty}$ have been so far performed only for powdered samples we introduce here the angle $\theta$ between the anisotropy axis of the Dy-ion and the magnetic field. 
 Although in
Ref. [\onlinecite{Dy2}] the authors used exact GCTM technique, they
found the eigenvalues of the transfer-matrix numerically. Thus, they
presented exact description of the thermodynamics of the model,
without any account of the ground stated properties. Investigating
the correspondence between theoretical and experimental
magnetization curves and susceptibility, they suggested the
following values of the parameters appearing in the
[DyCuMoCu]$_{\infty}$ material Hamiltonian\cite{Dy1,Dy2}
(\ref{ham}): $J_1=8.3$ cm$^{-1}$, $J_2=-11.8$ cm$^{-1}$, $K_1=-15.3$
cm$^{-1}$, and $K_2=8$ cm$^{-1}$. The values of the $g$-factors of
the magnetic ions are $g_1=2.16, g_2=1.97$ and $g_3=19.6$.
\begin{figure}
\includegraphics[width=\columnwidth]{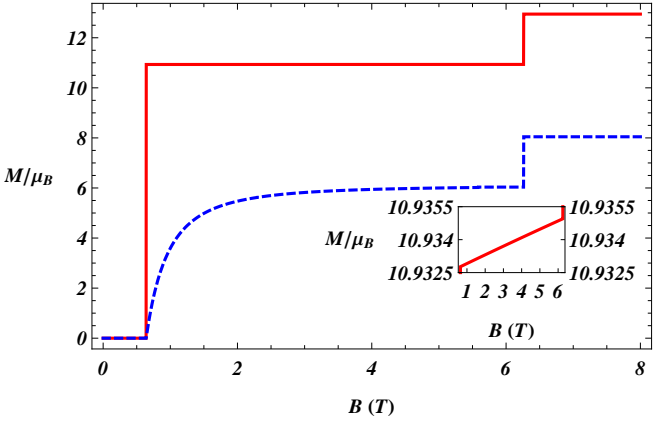}
\caption{\label{fig2}(color online) The plots of the magnetization
processed  $T=0.00001 
K$ for the model of a single-crystal [DyCuMoCu]$_{\infty}$(solid red) and powder sample (dashed blue). The inset shows weak but
monotonous growth of the magnetization at the broad
quasi-plateau between $B_1= 0.64093$ T with $M\approx
10.9325\mu_B$ and $B_2= 6.26021$ T with $M\approx 10.9350\mu_B$.
}
\end{figure}
\begin{figure}
\includegraphics[width=\columnwidth]{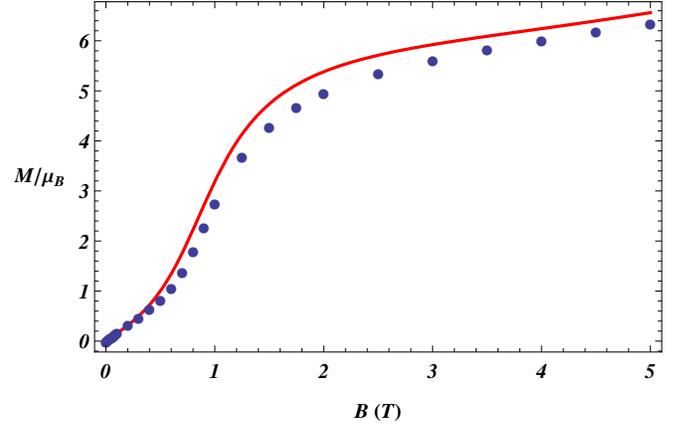}
\caption{\label{fig3n}(color online) Powder magnetization of [DyCuMoCu]$_{\infty}$ at 
$T=2K$ obtained by the angle integration of the theoretical curve (red solid) and experimental points (See Refs.[\onlinecite{Dy1,Dy2}])
}
\end{figure}
 The analytical treatment within the
framework of GCTM technique we are going to develop here not only
allows us to analyse the plots of the thermodynamic functions but
also offers an opportunity to identify all zero-temperature ground
states of [DyCuMoCu]$_{\infty}$ as well as to reveal the additional
physical feature of the magnetization process, the quasi-plateau.
 Let us start with the diagonalization of the block Hamiltonian
$\mathcal{H}_n$ which is the eight by eight matrix, having
block-diagonal form with two one-dimensional and two
three-dimensional blocks, corresponding to eigenstates of the Heisenberg trimer
with $S_{tot}^z=S_1^z+S_2^z+S_3^z=\pm 3/2$ and $\pm 1/2$
respectively. The eigenvalues and eigenstates corresponding to
$S_{tot}^z=\pm 3/2$ can be found easily:
\begin{eqnarray}\label{eps_1,8}
\varepsilon_{1}(\sigma_n,\sigma_{n+1})&=&\frac{1}{4}\left(
J_1+J_2\right)-\frac{1}{2}\left(2g_1+g_2\right)B\\
&+&\frac{1}{2}\left(K_1\left(\sigma_n+\sigma_{n+1}\right)+K_2\sigma_n\right)\nonumber\\
\varepsilon_{8}(\sigma_n,\sigma_{n+1})&=&\frac{1}{4}\left(
J_1+J_2\right)+\frac{1}{2}\left(2g_1+g_2\right)B\nonumber\\
&-&\frac{1}{2}\left(K_1\left(\sigma_n+\sigma_{n+1}\right)+K_2\sigma_n\right).\nonumber
\end{eqnarray}
For finding the rest six eigenvalues one has to consider a solution
of the cubic equations resulting from the eigenvalue problem for the
following matrices, corresponding to the $S_{tot}^z=\pm 1/2$
respectively:
\begin{eqnarray}\label{H1/2}
{\mathcal{H}}_{\pm 1/2}=\left( \begin{array}{ccc}
      h_{1}^{\pm}  & \frac{J_1}{2} & 0 \\
      \frac{J_1}{2}  &  h_2^{\pm} & \frac{J_2}{2}\\
      0  &  \frac{J_2}{2} & h_{3}^{\pm}
      \end{array}\right).
\end{eqnarray}
where we introduced the following notations:
\begin{eqnarray}\label{hhh}
h_1^{\pm}&=&\frac{1}{4}(J_2-J_1)\mp\frac{1}{2}\left(g_2 B+K_1
(\sigma_n-\sigma_{n+1})-K_2\sigma_n\right),\nonumber\\
h_2^{\pm}&=&\frac{1}{4}(J_1+J_2)\\
&\mp&\frac{1}{2}\left\{\left(2g_1-g_2\right)B-K_1 (\sigma_n+\sigma_{n+1})+K_2\sigma_n\right\},\nonumber \\\nonumber \\
h_3^{\pm}&=&\frac{1}{4}(J_1-J_2)\mp\frac{1}{2}\left(g_2 B-K_1
(\sigma_n-\sigma_{n+1})-K_2\sigma_n\right).\nonumber
\end{eqnarray}
 Thus, the rest six eigenvalues of the
Hamiltonian are given by the trigonometric solution of the cubic
equation,
\begin{eqnarray}\label{eps 27}
&&\varepsilon_j^{\pm}\left(\sigma_n,\sigma_{n+1}\right)=2\sqrt{Q_{\pm}}\cos\left(\frac{\phi_{\pm}+2\pi
j}{3}\right)-\frac{a^{\pm}_2}{3},\nonumber\\
&&j=0,1,2.
\end{eqnarray}
with
\begin{eqnarray}\label{QR}
&&\phi_{\pm}=\arccos\left(\frac{R_{\pm}}{\sqrt{Q_{\pm}^3}}\right), Q_{\pm}=\frac{(a^{\pm}_2)^2-3a^{\pm}_1}{9},\\
&&R_{\pm}=\frac{9a^{\pm}_1a^{\pm}_2-27a^{\pm}_0-2(a^{\pm}_2)^3}{54},\nonumber
\end{eqnarray}
and
\begin{eqnarray}\label{aaa}
&&a^{\pm}_2=-(h_1^{\pm}+h_2^{\pm}+h_3^{\pm}),\\
&&a^{\pm}_1=-\frac{1}{4}\left(J_1^2+J_2^2\right)+h_1^{\pm}h_2^{\pm}+h_1^{\pm}h_3^{\pm}+h_2^{\pm}h_3^{\pm},\nonumber\\
&&a^{\pm}_0=\frac{1}{4}\left(J_1^2 h_3^{\pm}+J_2^2
h_1^{\pm}\right)-h_1^{\pm}h_2^{\pm}h_3^{\pm}.\nonumber
\end{eqnarray}
\begin{figure}
\includegraphics[width=\columnwidth]{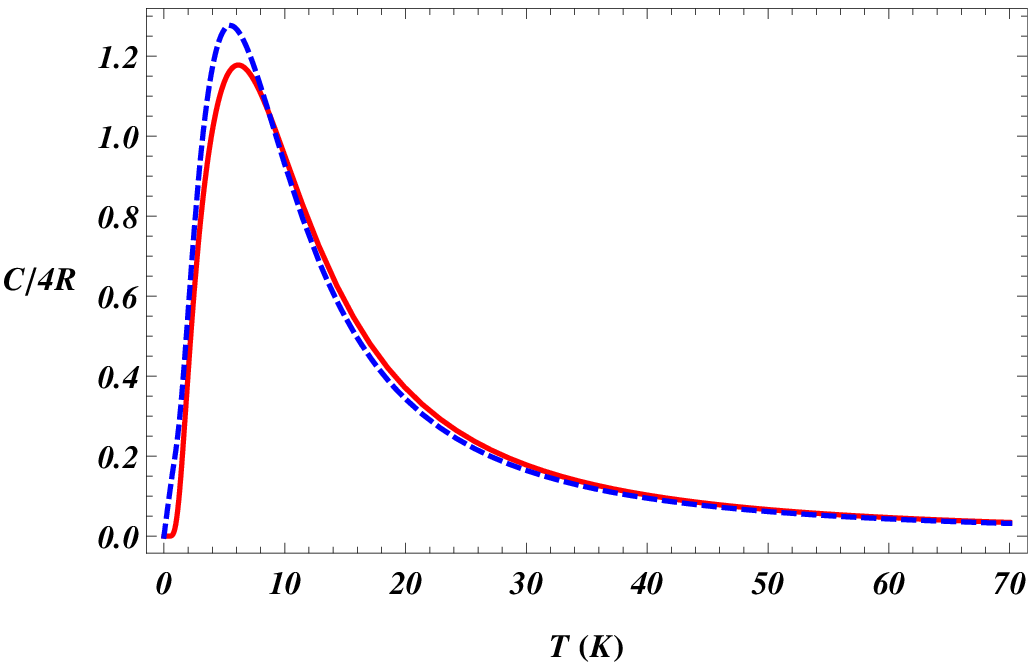}
\caption{\label{fig5n}(color online) Theoretical specific heat temperature dependence plots for the [DyCuMoCu]$_{\infty}$ at the value of magnetic field corresponding to the first 
quantum phase transition(level crossing) at $T=0$, $B=0.64093$ T. Solid red line corresponds to the sinlge crastall sample, dashed blue line - to powder. Here R is the gas constant. 
The factor $1/4$ appears because the within our formalism the free energy is given per one block containing four spins.}
\end{figure}
\begin{figure}
\includegraphics[width=\columnwidth]{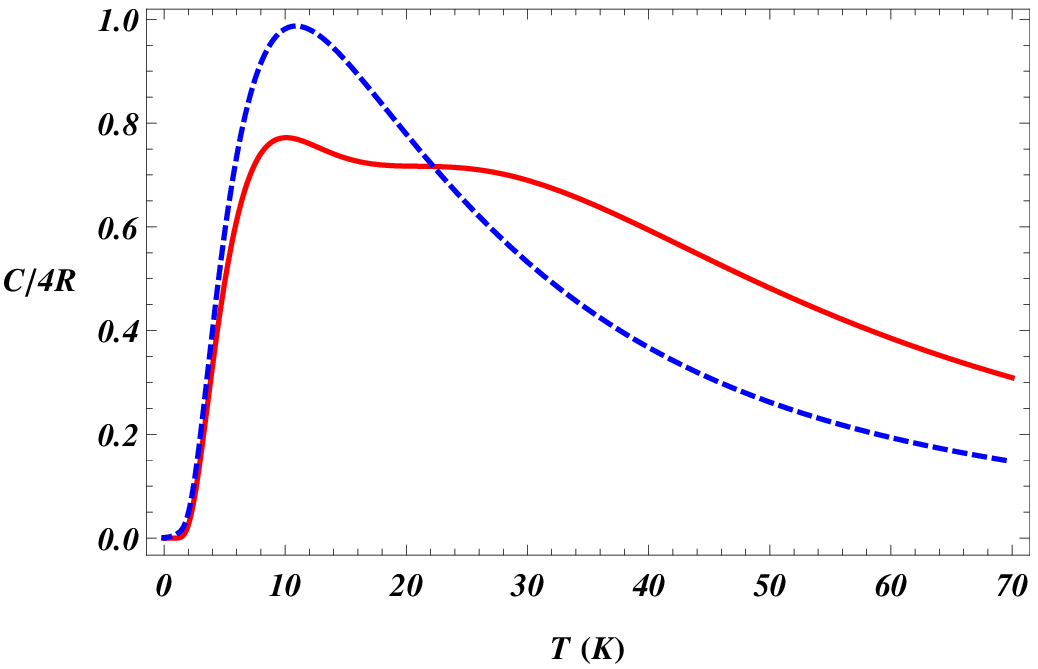}
\caption{\label{fig6n}(colour online) Theoretical specific heat temperature dependence plots for the [DyCuMoCu]$_{\infty}$ at the value of magnetic field corresponding to the second 
quantum phase transition(level crossing) at $T=0$, $B=6.26021$ T. Solid red line corresponds to the single crystal sample, dashed blue line - to powder.  Here R is the gas constant. 
The factor $1/4$ appears because the within our formalism the free energy is given per ona block containing four spins.
}
\end{figure}
Having all eight eigenvalues of the block Hamiltonian one can obtain
also the eigenstates corresponding to these eigenvalues:
\begin{eqnarray}\label{EV}
&&|3/2\rangle=|\uparrow\uparrow\uparrow\rangle\;\;\;\;\;\;\;\Rightarrow\;\;\varepsilon_1,\\
&&|-3/2\rangle=|\downarrow\downarrow\downarrow\rangle\;\;\;\Rightarrow\;\varepsilon_8,\nonumber\\
&&|\pm 1/2;j\rangle=\frac{A^{\pm}_j|\uparrow\uparrow\downarrow\rangle+|\uparrow\downarrow\uparrow\rangle+B^{\pm}_j|\downarrow\uparrow\uparrow\rangle}{\sqrt{1+(A^{\pm}_j)^2+(B^{\pm}_j)^2}}\nonumber\\
&&\Rightarrow\;\;\varepsilon^{\pm}_j,\;\; j=0,1,2,\nonumber
\end{eqnarray}
where
\begin{eqnarray}\label{AB}
A_j^{\pm}=\frac{J_{1}}{2 (\varepsilon_j^{\pm}- h_1^{\pm})},\;\quad
B_j^{\pm}=\frac{J_{2}}{2 (\varepsilon_j^{\pm}- h_3^{\pm})}.
\end{eqnarray}
In virtue of the presence of several g-factors the expectation value of the magnetic moment
corresponding to these ground states is non trivial. As
\begin{eqnarray}\label{m}
m^j=\mu_B g_1 \langle j|\left(S_1^z+S_3^z\right)|j\rangle+\mu_B g_2\langle
j|S_2^z|j\rangle,
\end{eqnarray}
 for the eigenstates with $S^z_{tot}=\pm 3/2$ one has just
$m_{\pm 3/2}=\pm\mu_B \left(2g_1+g_2\right)/2$. However, for the rest of the
eigenstates the magnetization depends on the system parameters
including magnetic field:
\begin{eqnarray}\label{m1/2}
m_{\pm1/2}^j=\pm\mu_B \frac{2g_1+g_2\left\{(A^{\pm}_j)^2+(B^{\pm}_j)^2-1\right\}}{2\left\{(A^{\pm}_j)^2+(B^{\pm}_j)^2+1\right\}},
\end{eqnarray}
where $j=0,1,2$. It is easy to see that at $g_1=g_2=g$ this
expression become equal to $\pm 1/2g\mu_B$. Having the exact eigenvalues
of the block Hamiltonian one can construct the GCTM for the whole
chain in the straightforward
way\cite{ant09,oha09,oha10,bel10,roj11a,oha12,bel12}:
\begin{eqnarray}\label{T}
&&T_{\sigma_n,\sigma_{n+1}}=e^{\frac{\beta \mu_Bg_3
B \cos \theta}{2}\left(\sigma_n+\sigma_{n+1}\right)}\mbox{Tr}\;e^{-\beta
\mathcal{H}_n}\\
&&=e^{\frac{\beta \mu_B g_3
B\cos \theta}{2}\left(\sigma_n+\sigma_{n+1}\right)}\sum_{\mbox{eigenvalues}}
e^{-\beta \varepsilon_j(\sigma_n,\sigma_{n+1})},\nonumber
\end{eqnarray}
where $\beta=1/k_B T$ is the inverse temperature. The partition
function for the whole chain under the periodic boundary conditions
is expressed in terms of the trace of the $N$-th power of the
transfer-matrix,
\begin{eqnarray}\label{Z}
Z=\sum_{\left(\sigma\right)}\prod_{n=1}^NT_{\sigma_n,\sigma_{n+1}}=\mbox{Tr}\;\mathbf{T}^N
\end{eqnarray}
 Thus, the free
energy per block in the thermodynamic limit is expressed through the
largest eigenvalue of the transfer-matrix in the following way:
\begin{widetext}
\begin{eqnarray}\label{f}
f(T,B,\theta)=-\frac{1}{\beta}\log\left[\frac{1}{2}\left(T_{\frac{1}{2},\frac{1}{2}}+T_{-\frac{1}{2},-\frac{1}{2}}
+\sqrt{\left(T_{\frac{1}{2},\frac{1}{2}}-T_{-\frac{1}{2},-\frac{1}{2}}\right)^2+4T_{\frac{1}{2},-\frac{1}{2}}T_{-\frac{1}{2},\frac{1}{2}}}\right)\right].
\end{eqnarray}
\end{widetext}
All thermodynamic functions, including magnetization, entropy,
specific heat, etc, can be obtained from the Eq. (\ref{f}) by means
of taking the corresponding derivatives with respect to $T$ and $H$. However, for the powdered sample one has to perform the angle averaging with respect to the mutual orientation of the
Dy ion anisotropy axis and magnetic field. Thus, for the magnetization and specific heat one will have
\begin{eqnarray}\label{MM}
&&M(T,B,\theta)=-\left(\frac{\partial f(T,B,\theta)}{\partial B}\right)_T,\\
&&M(T,B)=\int_0^{\pi/2}M(T,B,\theta)\sin \theta d \theta, \nonumber
\end{eqnarray}
\begin{eqnarray}\label{CC}
&&C(T,B,\theta)=-T\left(\frac{\partial^2 f(T,B,\theta)}{\partial T^2}\right)_B,\\
&&C(T,B)=\int_0^{\pi/2}C(T,B,\theta)\sin \theta d \theta. \nonumber
\end{eqnarray}

The plots of magnetization processes for both mono-crystalline and powdered [DyCuMoCu]$_{\infty}$ at very low temperature, $T=0.00001 K$ are presented in Fig. \ref{fig2}.
The comparison of the theoretical magnetization curve for powdered sample and the corresponding experimental points\cite{Dy2} for $T=2 K$ can be seen in Fig. \ref{fig3n}
The Fig. \ref{fig3n} completely reproduces the corresponding plots form Ref. [\onlinecite{Dy2}]. We include it here for completeness and for the consistency check for our 
analytic calculations. However, the analytic solution we obtained allows us not only to generate the plots of various thermodynamic quantities of the model but also 
gives us an advantage to describe the zero-temperature ground states corresponding to each part of the magnetization curve. This allows us, in its turn, to identify
the physical effect which has
not been revealed with the aid of numerical treatment of the GCTM
solution of the model in Ref. [\onlinecite{Dy2}], the quasi-plateau
structure in the magnetization curve (See inset of the Fig.
\ref{fig2}). The $S^z_{tot}=\pm1/2$ ground states of the spin trimer
Cu-Mo-Cu, in virtue of the difference between g-factors of Cu- and
Mo-ions, demonstrate monotonous dependence of the magnetization on
the external magnetic field. This feature is the reason of the quasi-plateau formation in the magnetization curve for the whole chain.
This is quite unusual phenomenon in the
physics of HIC. The typical $T=0$ (or for low enough temperatures)
magnetization curve of HIC consists of the horizontal parts
(magnetization plateaus) with a step-like transitions between them.
 This simple picture takes the place because of absence of
the excitation bands and quasi-particles with non-trivial dispersion
relation in HIC. On a contrary, the macroscopic ground states is
just a direct product of the local ground states of the building
blocks. Thus, the transitions between the magnetization plateaus
appear at the values of the magnetic field strength corresponding to
the level crossing of the spectrum of HIC. If the block Hamiltonian
of the HIC consists of only two quantum spins or if the g-factors
for all quantum spins are the same, the picture of exact
magnetization plateaus and step-like transitions between them is
always the
case\cite{Val,ant09,oha09,oha10,bel10,roj11a,RS11,SSR11,str12,oha12,bor_un,bel12,PSR,RSL,str05,str03,can06}.
However, as has been mentioned above, the magnetic properties of the HIC can be more rich and sophisticated. The reason for that, for example,
can be the eigenstates of the block Hamiltonian with the continuous dependence of the magnetic field, giving rise to the magnetization quasi-plateau
for the whole chain magnetization. The phenomenon of quasi-plateau caused by the difference in g-factors of the spins of block Hamiltonian of HIC has
been already mentioned in Ref. [\onlinecite{str05}], where the linear HIC
with quantum spin trimer has been proposed as the approximate model
for the F-F-AF-AF spin chain compound
Cu(3-Chloropyridine)$_2$(N$_3$)$_2$ (CCPA). There are three nonequivalent positions of the Cu$^{2+}$ ions in CCPA which lead to three different g-factors
but due to small difference between them ($g_1=2.1$, $g_2=2.14$, $g_3=2.19$)\cite{str05,CCPA} the effect is much less pronounces that in [DyCuMoCu]$_{\infty}$. In addition to that, CCPA hardly can
be considered as HIC. The HIC model of CCPA has been proposed in Ref. [\onlinecite{str05}] just as a very simplified exactly solvable model giving soem relevant results on magnetization and
susceptibility. It also demonstrate the quasi-plateau due to mechanism explained above. However, there is no systematic investigation of this issue from the experimental point of view. 
And even if CCPA would demonstrate the quasi-plateau phenomenon the mechanism apparently should be different form that inherent in HIC. 
The
case of the [DyCuMoCu]$_{\infty}$ , thus, can be considered as an
example of real HIC-material exhibiting quasi-plateaus according the
the mechanism describing above.
The values of the magnetization(scaled to the saturation magnetization),
corresponding to these local ground states of the quantum spin
trimer may be irrational (See Eq. (\ref{m1/2})). The appearance of
irrational magnetization plateaus does not contradict the
Oshikawa-Yamanaka-Affleck (OYA) criterion\cite{OYA}, as any
irrational values of magnetization, $2\sum_{a=1}^q g_a \langle
S_{a}^z\rangle/\sum_{a=1}^q g_a$, in the HIC correspond to the
expectation values of the spin operators which satisfy the OYA
criterion, $q(S-1/q \sum_{a=1}^q \langle S_{a}^z\rangle)$=integer.
Here $q$ is the spatial period of the ground state. It is worth
mentioning, that the irrational magnetization plateaus can also
appear in the integrable models of antiferromagnetic spin-S chains
doped with mobile spin-(S-1/2)carriers\cite{Frahm}. This fact is a
general feature of the HIC with the quantum trimers bearing
different g-factors on different spins\cite{bor_un}. Theoretical
magnetization curve of the [DyCuMoCu]$_{\infty}$ contains a plateau
at $M=0$ and a quasi-plateau starting at $M\approx10.9325$
and ending at $M\approx10.9350$. Thus, the deviation from the
ideal plateau is of the order $10^{-3}$. Constructing the whole set
of all possible zero-temperature ground states\cite{bor_un} and
comparing the internal energy obtaining from the free energy from
Eq. (\ref{f}) with the corresponding eigenvalues one can find the
macroscopic ground states corresponding to (quasi-)plateaux. The
zero-field ground state of the [DyCuMoCu]$_{\infty}$ compound is
found to be the antiferromagnetic one with doubling of the unit cell
and spatial modulation of the local magnetization in the
"up"-"down"-"down"-"up" form:
\begin{eqnarray}\label{AF}
|AF\rangle=\prod_{n=1}^{N/2}|\uparrow\rangle_{2n}|-1/2;1\rangle_{2n}|\downarrow\rangle_{2n+1}|1/2;1\rangle_{2n+1}.
\end{eqnarray}
Thus, the doubled unit cell contains eight magnetic ions, two
Dy$^{3+}$ ions spin pointing oppositely to each other and two
composite $S=1/2$ spins of the Cu-Mo-Cu quantum spin trimers 
pointing again oppositely to each other and oppositely to the Dy$^{3+}$
ion spin from the left hand site, thus, parallel to the Dy$^{3+}$
spin from the right hand site. Thus, there are two
antiferromagnetically ordered sublattices of the system: the spins
of Dy$^{3+}$ ions form a N\'{e}el ordered structure, and so do the
composite $S=1/2$ spins of the quantum trimers. Moreover, inside
each block spins of Dy$^{3+}$ and of quantum trimers are
antiparallel. The quantum phase transition (at zero temperature)
which takes place at $B\approx 0.64093$ T suddenly rises the system
magnetization to the value $M\approx 10.9325$. Under the
action of the external magnetic field Ising spins which were
pointing down, as well as the composite $S=1/2$ spins of the quantum
spin trimers which were pointing down undergo a flip, however, the
composite structure of the local quantum state of the Cu-Mo-Cu spin
trimers does not change. As a result, magnetization undergoes a jump
and the corresponding zero-temperature ground state become the
following:
\begin{eqnarray}\label{QP}
|QP\rangle=\prod_{n=1}^N|\uparrow\rangle_n|1/2;1\rangle_n.
\end{eqnarray}
The magnetization of this quasi-plateau state has very weak but
monotonous dependence of the external magnetic field due to Eq.
(\ref{m1/2}) and is given by
\begin{eqnarray}\label{MQP}
M_{QP}=\frac{\mu_B}{2}\left(g_3+\frac{2g_1+g_2
\left\{(A^{+}_1)^2+(B^{+}_1)^2-1\right\}}{(A^{+}_1)^2+(B^{+}_1)^2+1}\right)
\end{eqnarray}
\begin{figure}[h]
\begin{center}
\includegraphics[width=\columnwidth]{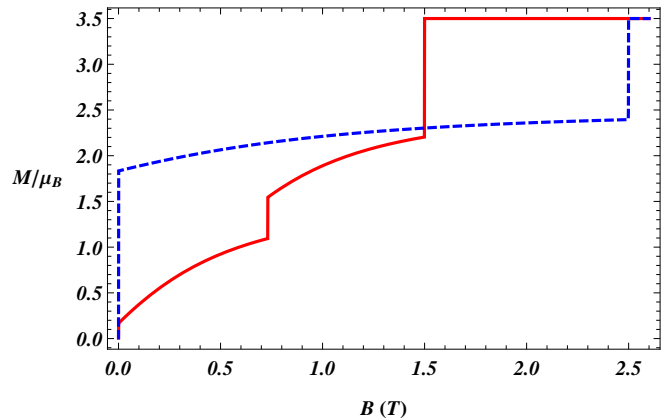}
\caption{\label{fig6} (Color online) Low temperature magnetization
curves for the isolated spin-1/2 trimer with the structure of
Cu-Mo-Cu part of the [DyCuMoCu]$_{\infty}$. The solid red line
corresponds to $J_1=J_2=1$, $g_1=1$, $g_2=5$; dashed blue line
corresponds to $J_1=J_2=1$, $g_1=3$, $g_2=1$; $T=0.00001$ K. }
\end{center}
\end{figure}
If the g-factors of Cu and Mo would be the same ($g_1=g_2=g$), the
magnetization corresponding to the $|QP\rangle$ state would no
longer be a monotonous function of the external magnetic field, but
just a constant, $\mu_B\frac{g+g_3}{3 g+g_3}$, which takes the value
$1/2\mu_Bg$ at $g_3=g$. The phenomenon of quasi-plateau is a direct
consequence of the appearance of trimers of quantum spins with
different g-factors and a separable character of the macroscopic
ground state for the whole chain which, it its turn, stems from the
fact that the material under consideration is an example of HIC.
It is also important to figure out how one can identify the effect of quasi-plateau and to distinguish it form the 
the growth of magnetization on the real plateau due to thermal fluctuations. The issue in general is complicated and is highly affected by many facts, particularly by the
character of the eigenstate magnetic moment dependence on the magnetic filed and temperature. Here,  we would like just to estimate roughly the relative growth of magnetization
caused by the thermal fluctuation (``slope'' of the plateau) and by the quasi-plateau effect for  the model of monocrystalline [DyCuMoCu]$_{\infty}$ at $T=0.00001 K$(Fig. \ref{fig2}).
For the real plateau at $M=0$ one gets $\Delta M /\Delta B\approx 0.00009$, while for the quasi-plateau $\Delta M /\Delta B\approx 0.00056$ which includes of cause the thermal effects as well. So, for the 
given material at least for low enough temperature the quasi-plateau slope is of one order larger that just the thermal one.
Though, in [DyCuMoCu]$_{\infty}$ the observing deviation from ideal
plateau of the part of magnetization curve corresponding to
$|QP\rangle$ ground state is very small, it is the quantitative
effect and with the change of values of coupling constants and/or
g-factor it can be pronounces in a most prominent way, even
noticeable to a naked eye\cite{bor_un}. One of the crucial
circumstances for that is the difference between the g-factors of
spins. To illustrate this effect the almost zero temperature
($T=0.00001$ K) magnetization curves for the isolated spin trimers
with the same structure as Cu-Mo-Cu in is shown in Fig. \ref{fig6}. One can see the
well pronounced quasi-plateaus with non-linear dependence on $B$ for
the case $g_1=1$, $g_2=5$ as well as for the case $g_1=3$, $g_2=1$.
Thus, in order to get a HIC material with well pronounced
magnetization quasi-plateau according to the mechanism described in
the present letter the further search for the single-chain magnets
containing highly anisotropic ions and polymetallic complexes with
the large difference between the g-factor should be conducted.
Another important possibility to enhance the deviation  of the
quasi-plateaus from the horizontal line is the electric field
control of g-factor\cite{g1,g2}.
In Figs. \ref{fig5n} and \ref{fig6n} we presented the plots of the thermal behavior of specific heat for two critical values of $B$.The plots are obtained only within the model considered here,
so for the comparison with experimental results the corrections form the excited Kramers doublets of Dy$^{3+}$ is needed, which is out of scope of our paper. However, one can see how the powder
averaging increases the specific heat at low temperatures and decreases it when the temperature is high.

 In summary, we have constructed the exact analytical description of
 the magneto-thermal properties of the SCM with Ising and Heisenberg
 bonds,[DyCuMoCu]$_{\infty}$. We have identified the
 zero-temperature ground states of the material corresponding to
 the magnetization curve. From $B=0$ T to $B\approx0.64093$ T the
 material is in the antiferromagnetic ground state with spatially modulated spin
 order with two antiferromagnetically ordered sublattices of
 Dy$^{3+}$ ions and Cu-Mo-Cu trimers which have $S=1/2$. The
 sublattices have also antiferromagnetic order with respect to each
 other, forming the spatial modulation of the local $S=1/2$ spin
 states in the form "up"-"down"-"down"-"up". From $B\approx0.64093$ to
 $B\approx6.26021$ the ground state differs from the previous one by
 the flipping of all Dy$^{3+}$ spins as well as Cu-Mo-Cu trimer
 $S=1/2$ spins up. As the magnetization of the corresponding ground
 state depends monotonically on the magnetic field the corresponding
 part of the magnetization is a quasi-plateau with the explicit 
 dependence on $B$.

 The authors express their deep gratitude to Willem Van den Heuvel for shearing the experimental data and for
  very helpful communication. We are thankful to Temo Vekua for stimulating discussions.
  V. O. express his gratitude to the LNF-INFN for warm hospitality
  during the work on the letter. He also acknowledeges the partial financial from the 
  grant of the State Committee of Science of Armenia No. 13-1F343. 


\begin{thebibliography}{99}
\bibitem{footnote1} Here we consider the simplified model of the magentic lattice of [DyCuMoCu]$_{\infty}$ proposed in Ref. [\onlinecite{Dy2}]. 
Actually, two different Cu$^{2+}$ ions ($\mathbf{S}_{n,1}$ and $\mathbf{S}_{n,3}$) have not only different g-factors ($g_x=2.063$, $g_y=2.079$, $g_z=2.328$ for $\mathbf{S}_{n,1}$ 
and $g_x=2.063$, $g_y=2.084$, $g_z=2.347$ for $\mathbf{S}_{n,3}$) but they are quite anisotropic. The same is valid for the Mo$^{5+}$ ion. Thus, the more acurate model of [DyCuMoCu]$_{\infty}$ 
must not only take into account different g-factors for $\mathbf{S}_{n,1}$ and $\mathbf{S}_{n,3}$ but also should contain four different couplig constants.
\end{thebibliography}
\end{document}